\documentclass{article}
\usepackage{graphicx,emulateapj}

\begin{document}

\title{The Intracluster Medium in $z > 1$ Galaxy Clusters}

\author{S.A.\ Stanford\altaffilmark{1}, Bradford Holden\altaffilmark{1}}

\affil{Physics Department, University of California-Davis, Davis, CA 
95616, USA}
\authoremail{adam,bholden@igpp.ucllnl.org}
\altaffiltext{1}{Institute of Geophysics and Planetary Physics,
Lawrence Livermore National Laboratory}

\author{Piero Rosati}
\affil{European Southern Observatory, Karl-Scharzschild-Strasse 2,
D-85748 Garching, Germany}
\authoremail{prosati@eso.org}

\author{Paolo Tozzi, Stefano Borgani}
\affil{Osservatorio Astronomico di Trieste, via G.B.\ Tiepolo 11,
I-34131, Trieste, Italy}
\authoremail{tozzi,borgani@ts.astro.it}

\author{Peter R. Eisenhardt}
\affil{Jet Propulsion Laboratory, California Institute of Technology,
MS 169-327, 4800 Oak Grove, Pasadena, CA 91109}
\authoremail{prme@kromos.jpl.nasa.gov}

\and 

\author{Hyron Spinrad}
\affil{Astronomy Department, University of California, Berkeley, CA
94720}
\authoremail{spinrad@bigz.berkeley.edu}

\begin{abstract}
The Chandra X-ray Observatory was used to obtain a 190 ks image of
three high redshift galaxy clusters in one observation.  The results
of our analysis of these data are reported for the two $z > 1$
clusters in this Lynx field, which are the most distant known X-ray
luminous clusters.  Spatially-extended X-ray emission was detected from
both these clusters, indicating the presence of hot gas in their
intracluster media.  A fit to the X-ray spectrum of RX~J0849+4452, at
$z=1.26$, yields a temperature of $kT = 5.8^{+2.8}_{-1.7}$ keV.  Using
this temperature and the assumption of an isothermal sphere, the total
mass of RX~J0849+4452 is found to be $4.0^{+2.4}_{-1.9} \times 10^{14}
h_{65}^{-1} M_{\sun}$ within $r = 1 h_{65}^{-1}$ Mpc.  The $T_x$ for
RX~J0849+4452 approximately agrees with the expectation based on its
$L_{bol} = 3.3^{+0.9}_{-0.5} \times 10^{44}$ erg s$^{-1}$ according to
the low redshift $L_x - T_x$ relation.  The very different
distributions of X-ray emitting gas and of the red member galaxies in
the two $z > 1$ clusters, in contrast to the similarity of the
optical/IR colors of those galaxies, suggests that the early-type
galaxies mostly formed before their host clusters.

\end{abstract}


\section{Introduction}

Numerical simulations based on hierarchical clustering models such as cold dark
matter (CDM) show the rate at which clusters form depends critically on
$\Omega_m$, and weakly on $\Lambda$ and the initial power spectrum.  Thus, the
observed evolution of the cluster number density of a given X-ray temperature
and luminosity can determine $\Omega_m$ \cite{ob92,eke98a}.  A critical
component of such a measurement is an understanding of the thermodynamical
evolution of the intracluster medium (ICM) \cite{bower97,borg99,tn00}.

The study of ICM properties at high $z$ is mostly unexplored territory.  Until
the advent of Chandra and XMM, X-ray observations with limited resolution and
sensitivity were unable to provide conclusive measurements of the ICM at $z>1$
associated with bona fide galaxy clusters.  For example, ROSAT observations of
high $z$ radio galaxies possibly associated with galaxy clusters
\cite{crawfab96} provided only limited evidence that the observed X-ray
emission originates from hot intracluster gas.

To assemble a well--defined sample for studying the properties of the ICM and
cluster galaxy populations at $z > 1$, we have been using near-IR imaging and
deep optical spectroscopy to complete the identification of the faintest
candidate clusters in the Rosat Deep Cluster Survey (RDCS)
\cite{rosati95,rosati98}.  Of the 4 RDCS clusters which have been identified at $z
> 1$, two are separated by only 4.2 arcmin on the sky and 0.01 in redshift:
RX~J0848+4453 at $z = 1.27$ \cite{sas97} and RX~J0849+4452 at $z = 1.26$
\cite{rosati99}.  Keck LRIS spectroscopy has confirmed 20 member galaxies in
these two clusters.  In addition, a third moderate redshift cluster in this Lynx
field already had been identified from the RDCS, RX~J0848+4456 at $z=0.56$
(Rosati et al.\ 1998; the Chandra results on this object will be presented
elsewhere: Holden et al., in preparation).  The Lynx field offers the
opportunity to probe the physical parameters of the ICM over the redshift range
0.6--1.3 in a single deep Chandra pointing.  $H_0 = 65$ km s$^{-1}$ Mpc$^{-1}$,
$\Omega_m=0.3$, and $\Lambda=0.7$ are assumed throughout this paper.

\section{Observations and Reductions}

New Lynx field X-ray data were obtained by Chandra using the Advanced CCD
Imaging Spectrometer imaging (ACIS-I) detector.  Two exposures were obtained in
the faint mode when ACIS was at a temperature of $-$120 K.  The first
observation (Obs ID 1708) was taken on 2000 May 3 for 65 ks, and the second (Obs
ID 927) on 2000 May 4 for 125 ks.

The Level 1 data were processed using the calibration files available as of 15
September 2000 for the aspect solution and the quantum efficiency uniformity,
which correct the effective area for loss due to charge transfer inefficiency.
The data were filtered to include only the standard event grades 0, 2, 3, 4, and
6.  All hot pixels and columns were removed, as were the columns close to the
border of each node, since the grade filtering is not efficient in these
columns.  The removal of columns and pixels slightly reduces the effective area
of the detector, the effect of which has been included when calculating the
total exposure maps.  Time intervals with background rates larger than $ 3
\sigma$ over the quiescent value ($0.30$ counts s$^{-1}$ per chip in the 0.3 -
10 keV band) were removed.  This procedure gave 61 ks of effective exposure in
the first observation, and 124 ks in the second, for a total of 185 ks.

\section{Results}

Both $z > 1$ clusters were detected and found to contain spatially
extended X-ray emission (left panel of Figure~\ref{acisprof}).  The
emission in RX~J0848+4453 is weak and amorphous, while in
RX~J0849+4452 it is symmetric and centrally concentrated in a way
similar to that seen in relaxed clusters at lower redshift. The radial
surface brightness profile for RX~J0849+4452 is shown in the right
panel of Figure~\ref{acisprof} along with a $\beta$ model fit.  In
both clusters there are relatively strong point sources within the
{\tt ROSAT} detection area.  In both cases, approximately 40\% of the
flux as measured by {\tt ROSAT} is from point sources unlikely to be
associated with the clusters, showing the importance of obtaining high
spatial resolution data.  Further analyses of the point sources found
in the Chandra observation, including their optical-IR
identifications, will be presented by Stern et al.\ (in preparation).

The spectra of the two clusters were determined in the following way.
After removing the point sources, events at 0.5---6.0 keV in an
$r=35\arcsec$ ($\sim$0.32 Mpc) circular aperture centered on each
cluster were extracted.  Events in a surrounding background region,
with the cluster and point sources excised, were also extracted.  The
background model consists of a broken power law and a Gaussian for the
instrumental Au emission line at 2.1 keV.  The background models for
both clusters were fit separately and agree with each other within the
calculated errors.  The normalization of these background models was
rescaled by the relative areas of the background region and the
cluster apertures.  The number of net counts in the 0.5--6.0 keV range
is 182 and 430 for RX~J0848+4453 and RX~J0849+4452, respectively,
within the $r=35\arcsec$ aperture after the removal of point sources
and the background.  Figure~\ref{spectra} shows the unfolded spectra
(see below), which were divided into bins of 40 events before
background subtraction.

To determine the X-ray temperatures and fluxes, Raymond-Smith (1977) model
spectra with absorption were folded through the appropriate response matrices
and fit to the data, while the background model was held fixed, using a
maximum-likelihood test with the statistic from Cash (1979). 
The column density was fixed at $n(H) = 2.0 \times 10^{20}$ cm$^{-2}$, which
agrees fairly well with both the $n(H)$ obtained from the 100\micron\ maps of
Schlegel et al.\ (1998), and that derived using Dickey \& Lockman (1990).  The
metallicity of the model spectrum was fixed at 0.3 solar; changing the abundance
does not change the best fit temperature.  The best fit $T_x$ are $kT =
5.8^{+2.6}_{-1.7}$ and $1.6^{+0.8}_{-0.6}$ keV (one $\sigma$ uncertainties) for
RX~J0849+4452 and RX~J0848+4453, respectively.  These temperatures are in line
with the expectations based on the clusters' $L_x$ (Table \ref{tab1}) and the
low redshift $L_x - T_x$ relation.  This extends to $z \sim 1.3$ the results
found up to $z \sim 0.8$ by Donahue et al.\ (1999) and Della Ceca et al.\
(2000).  The apparent lack of evolution of the $L_x - T_x$ relation to such high
redshifts has important implications for both ICM thermodynamics and
cosmological models \cite{borg00}.

The spectral analysis gives the following results.  Within the $r =
35\arcsec$ measurement aperture, the observed frame $F_{0.5-6.0} =
1.4^{+0.2}_{-0.2} \times 10^{-14}$ and $0.44^{+0.08}_{-0.08} \times
10^{-14}$ ergs cm$^{-2}$ s$^{-1}$ for RX~J0849+4452 and RX~J0848+4453,
respectively.  The aperture fluxes and luminosities in the rest frame
0.5--2.0 keV band are given in Table~\ref{tab1}.  The total flux for
RX~J0849+4452, obtained by extrapolating the best fit King profile, is
$F_{0.5-2.0}^{rest} = 1.5^{+0.2}_{-0.2} \times 10^{-14}$ ergs
cm$^{-2}$ s$^{-1}$.

The uncertainties in the quantities presented above were estimated from
10$^4$ Monte Carlo simulations of the spectra.  The simulations are based on the
input models for the background and the cluster spectra.  The models are folded
through the appropriate response matrices and then randomly sampled.  The
simulated output is fit in the same way as the original spectra.  The resulting
distribution of temperatures is asymmetric but the median temperature matches
the input temperature, showing no strong biases in our method of measuring
$T_x$.

A total mass can be estimated from the $T_x$, assuming an isothermal
sphere and extrapolating the X-ray emission to $r = 1$ Mpc using the
best fit profile shown in the right panel of Figure~\ref{acisprof}.
The total mass of RX~J0849+4452 derived from the new X-ray data is
$4.0^{+2.4}_{-1.9} \times 10^{14} M_{\sun}$ within $r = 1 h_{65}^{-1}$
Mpc.  No estimate was calculated for RX~J0848+4453 due to its strong
departure from spherical symmetry and uncertain $T_x$.  The new $L_x$
for RX~J0848+4453 is approximately in line with the expectation
according to the present epoch $L_x - \sigma$ relation (e.g.\ Edge \&
Stewart 1991; Borgani et al.\ 1999b) for the estimate we have made of
the velocity dispersion $\sigma = 650 \pm 170$ km s$^{-1}$ of its 9
member galaxies for which we have sufficiently high resolution
spectra.  Member galaxy redshifts have not yet been obtained with
sufficient accuracy to enable the calculation of a meaningful velocity
dispersion for comparison to the total mass estimate in the case of
RX~J0849+4452.

\section{Discussion}

The spatially extended X-ray emission seen in RX~J0849+4452 clearly shows that a
hot intracluster medium exists in galaxy clusters at $z > 1$.  In this cluster
the ICM has a regular, centrally concentrated spatial distribution, similar to
the relaxed appearance RX~J0849+4452 presents in the optical/IR
(Figure~\ref{oix}).  But the ICM in the other $z > 1$ cluster, RX~J0848+4453, is
very weak and appears to be divided between the two sides of the cluster.  Two
groups may be in the process of merging, as appears to be seen also in the
distribution of the red galaxies (Figure~\ref{oix}).  Using photometric
redshifts for all galaxies at $K < 20.6$, nearly 50\% of the member galaxies lie
within the central 30 arcsec of RX~J0849+4452, while only 10\% of the probable
members are in the same region of RX~J0848+4453 (van Dokkum et al., in
preparation).  While the two clusters show rather different distributions both
in the optical/IR and in the X-ray, they contain galaxy populations dominated by
red galaxies which appear to have little or no recent star formation
\cite{sas97,rosati99}.  The median colors of these red galaxies are the same
within 0.05 in e.g.\ $I-K$, and HST imaging of RX~J0848+4453 shows them to be
overwhelmingly early-types (van Dokkum et al.\ in preparation).  This suggests
that the early-type galaxies found in clusters were largely formed prior to
cluster formation.

\medskip

We thank Scott Wolk for assistance with planning our Chandra
observation, and Patrick Wojdowski for help with Chandra data
analysis, and the referee for a timely report.  Support for SAS came
from NASA/LTSA grant NAG5-8430 and for BH from NASA/Chandra GO0-1082A,
and both are supported by the Institute of Geophysics and Planetary
Physics (operated under the auspices of the US Department of Energy by
the University of California Lawrence Livermore National Laboratory
under contract W-7405-Eng-48).  Portions of this work were carried out
by the Jet Propulsion Laboratory, California Institute of Technology,
under a contract with NASA.

\begin{deluxetable}{llccccc}
\tablewidth{0pt} 
\tablecaption{Chandra Results\tablenotemark{a}} 
\tablehead{ 
\colhead{Cluster} & 
\colhead{$z$} & 
\colhead{F[0.5-2.0]\tablenotemark{b}} &  
\colhead{L[0.5-2.0]\tablenotemark{b}} & 
\colhead{L[bol]\tablenotemark{c}} & 
\colhead{$T_x$} & 
\colhead{M$_{\rm total}$($<1$ Mpc)} \\
\colhead{} & 
\colhead{} & 
\colhead{$10^{-15}$ erg s$^{-1}$ cm$^{-2}$} &
\colhead{$10^{44}$ erg s$^{-1}$} &
\colhead{$10^{44}$ erg s$^{-1}$} & 
\colhead{keV} & 
\colhead{$10^{14}$ M${\sun}$}
} 
\startdata 
RXJ0848+4453 & 1.27 & 1.4$^{+0.3}_{-0.3}$ & 0.40$^{+0.16}_{-0.11}$ & 0.69$^{+0.27}_{-0.17}$ & 1.6$^{+0.8}_{-0.6}$ & \nodata \\[0.2cm]
RXJ0849+4452 & 1.26 & 9.0$^{+0.9}_{-0.9}$ & 0.51$^{+0.05}_{-0.05}$ & 3.3$^{+0.9}_{-0.5}$ & 5.8$^{+2.8}_{-1.7}$ & 4.0$^{+2.4}_{-1.9}$ \\
\enddata
\label{tab1}
\tablenotetext{a}{$H_0 = 65$ km s$^{-1}$ Mpc$^{-1}$,
$\Omega_m=0.3$, and $\Lambda=0.7$ are assumed.}
\tablenotetext{b}{$r=35\arcsec$ aperture}
\tablenotetext{c}{aperture corrected to total}
\end{deluxetable}

\newpage

\epsscale{0.7}
\plottwo{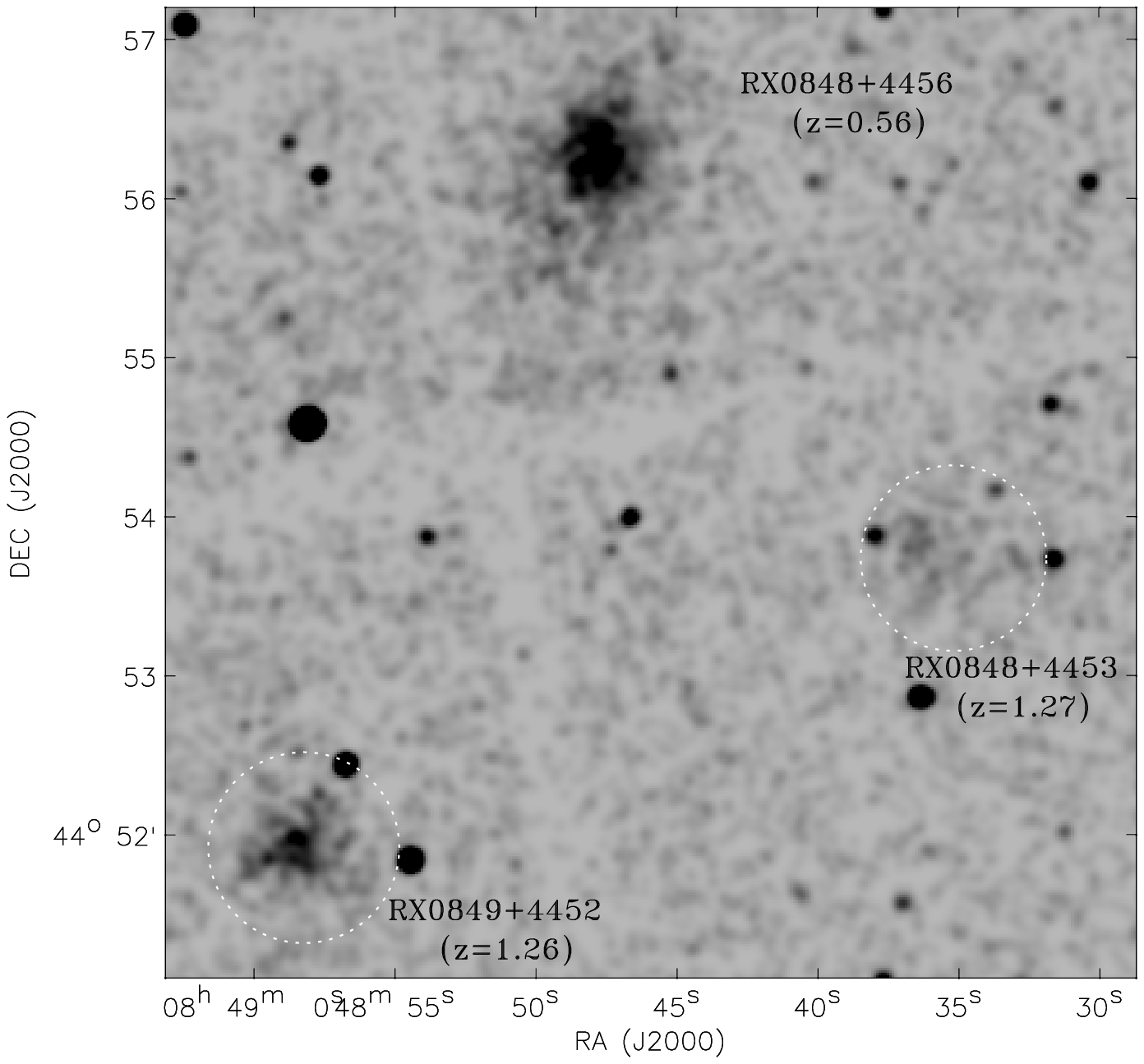}{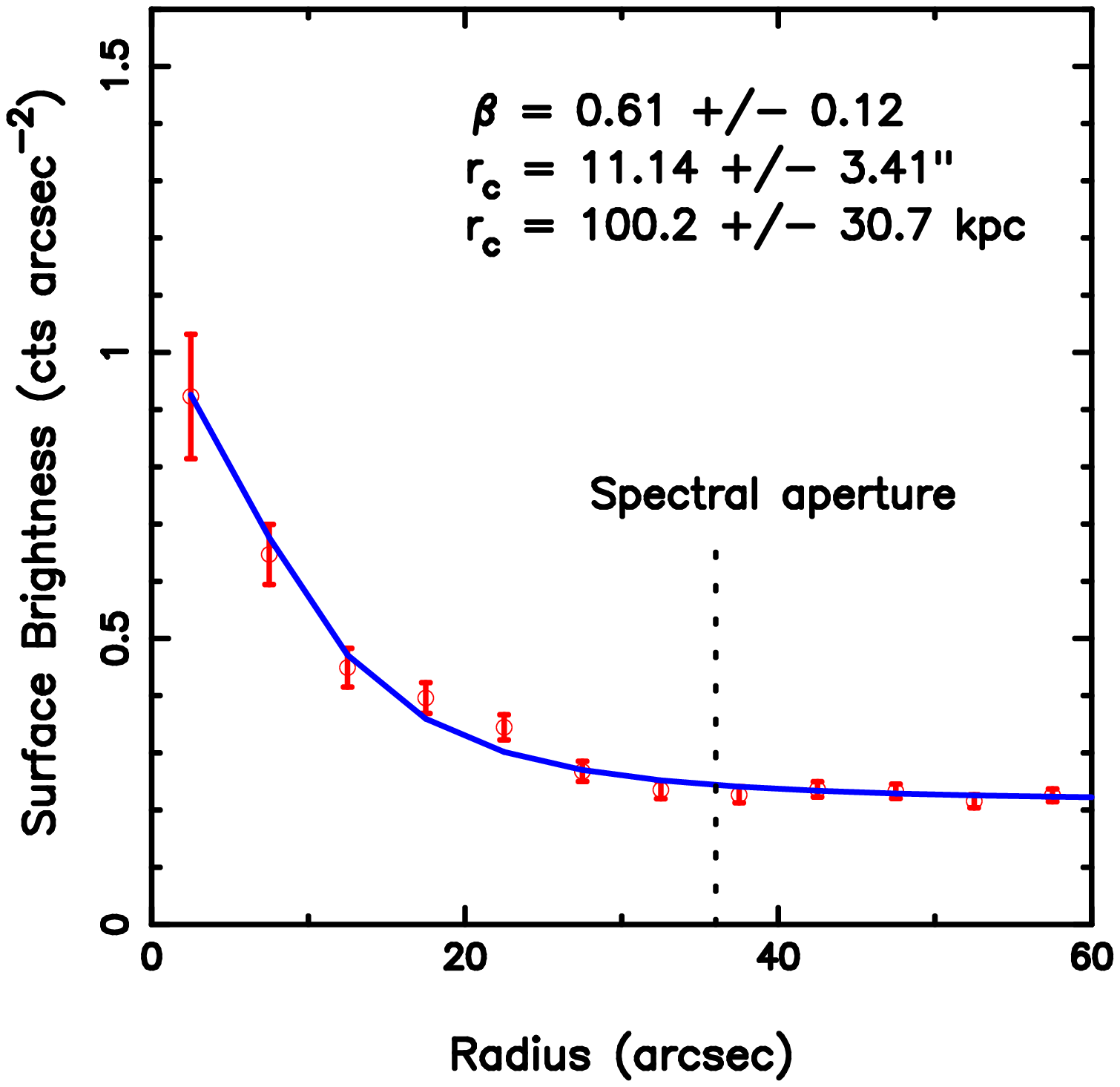}
\figcaption{(left) Greyscale image of the ACIS-I data showing a $6 \times 6$ 
arcmin field.  The image has been smoothed by a two-dimensional Gaussian with
$\sigma = 2.1$ arcsec. The apertures used for making spectra are shown by dashed
circles around the two higher-$z$ clusters; the point sources in these apertures 
were excluded when creating the spectra. (right) Radial profile of the X-ray
emission centered on RX~J0849+4452 shown by the small circles with one $\sigma$
errorbars.  A $\beta$ model fit is shown by the line, and the best fit
parameters are indicated.
\label{acisprof}
}

\bigskip

\epsscale{0.9}
\plottwo{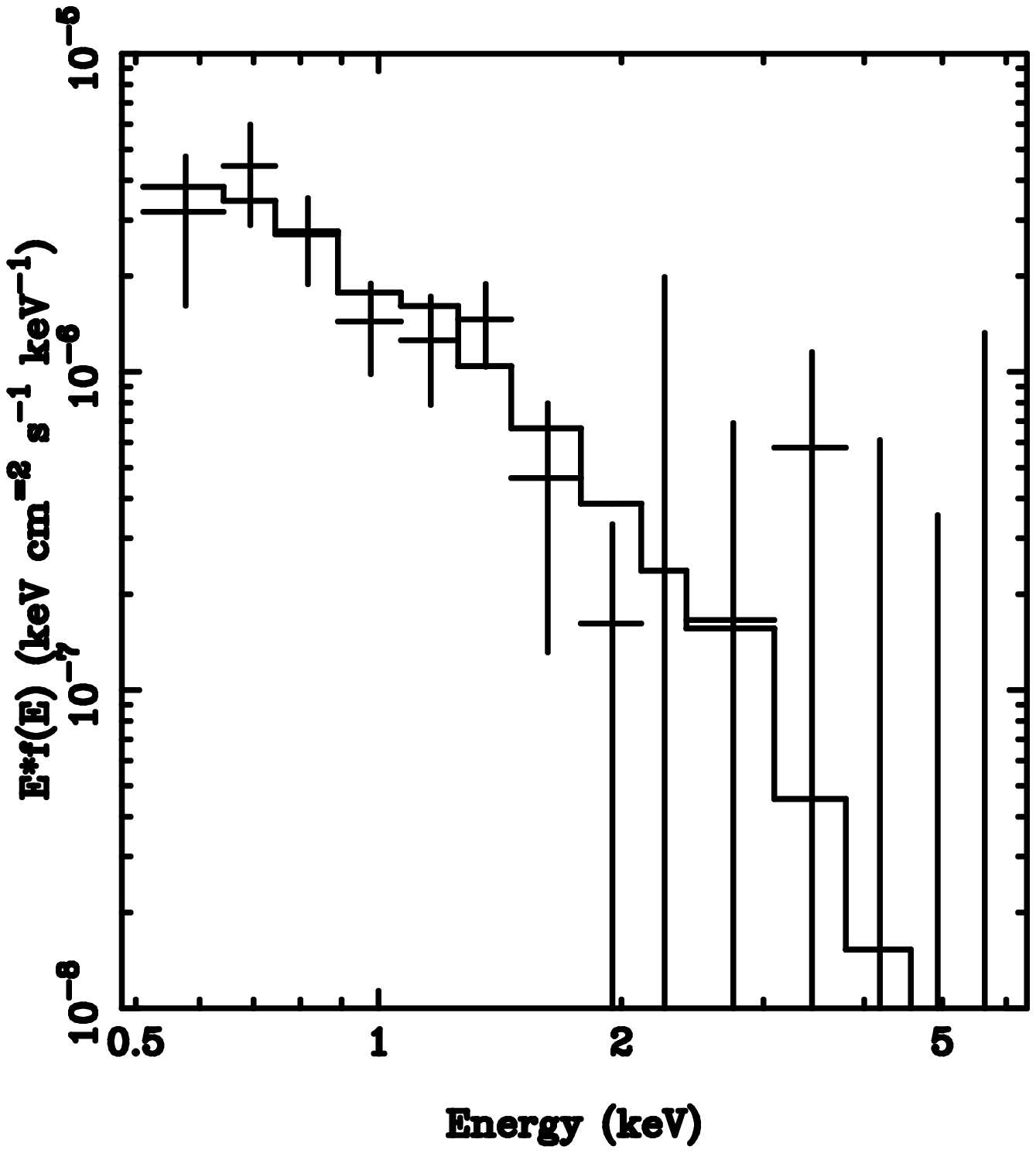}{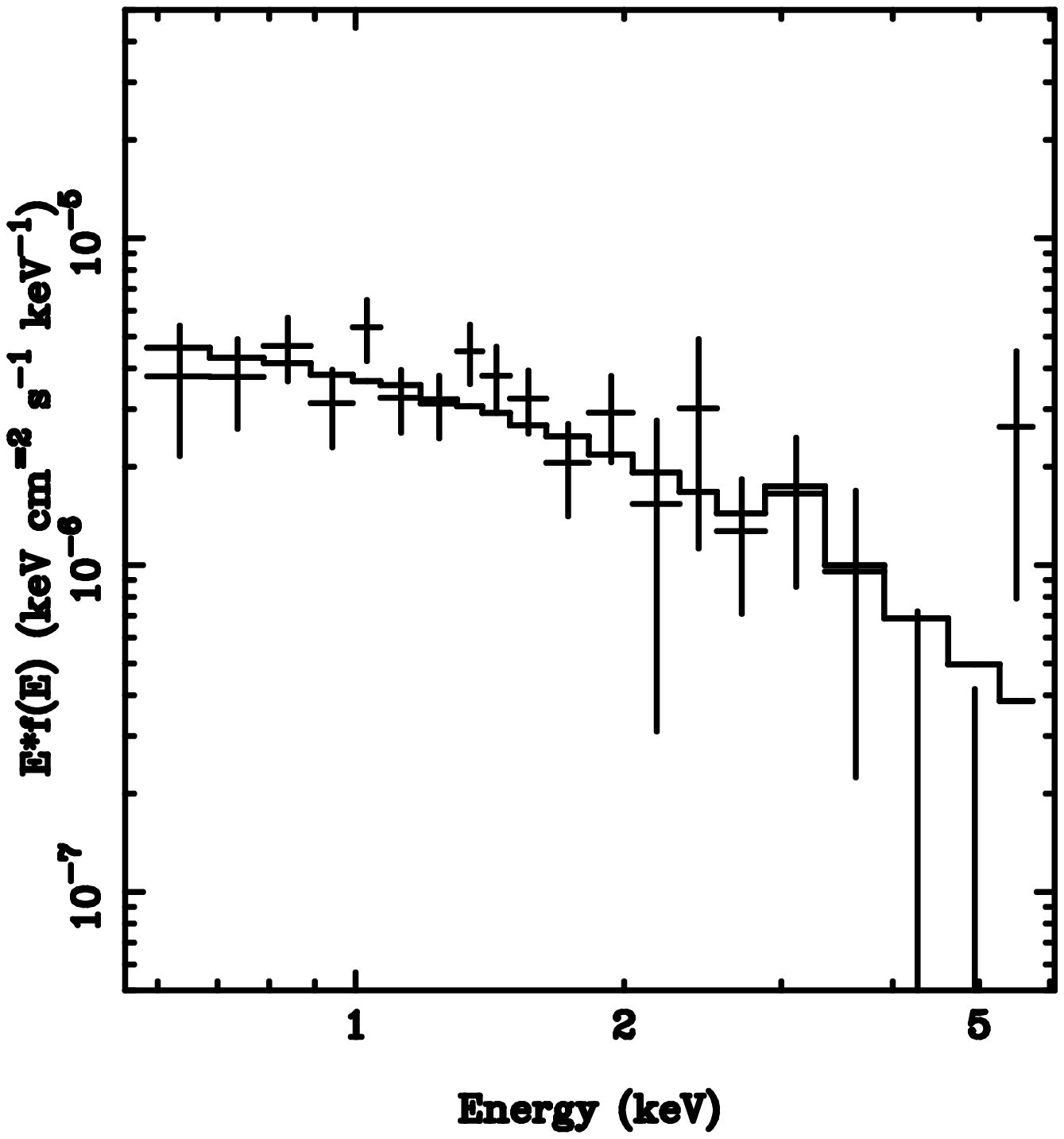}
\figcaption{ACIS-I unfolded spectra of RX~0848+4453 at $z=1.27$ (left) and 
RX~J0849+4452 (right; $z=1.26$) with the fitted models.
\label{spectra}
}

\bigskip

\epsscale{1.1}
\figcaption{$BIK$ images \cite{sas97,rosati99} showing a $2 \times 2$ 
arcmin field on RX~J0848+4453 (left; $z=1.27$) and RX~J0849+4452 (right; $z=1.26$)
with contour overlays (at the levels of $2.5,5,7,15 \sigma$) of the X-ray
emission detected by Chandra/ACIS-I.  The strong X-ray point sources centered on
optical objects in the vicinity of the cluster cores are unlikely to be
associated with the clusters.  In RX~J0848+4453 (left), the relatively bright
source at lower left, which is associated with a red galaxy, is a known quasar at
$z=1.194$.  In RX~J0849+4452 (right), the relatively bright X-ray source to the
north of the cluster core is associated with a quasar at
$z=1.329$.
\label{oix}
}


\begin{thebibliography}{}

\bibitem[Borgani et al.\ 1999]{borg99} Borgani, S., Rosati, P., Tozzi,
P., \& Norman, C. 1999, ApJ, 517, 40

\bibitem[Borgani et al.\ 1999]{borg99b} Borgani, S., Girardi, M.,
Carlberg, R.G., Yee, H.K.C., \& Ellingson, E.\ 1999, ApJ, 527, 561

\bibitem[Borgani et al.\ 2000]{borg00} Borgani, S., Rosati, P., Tozzi, 
P., Stanford, S.A., Eisenhardt, P.R., Lidman, C., Holden, B., Norman,
C., \& Squires, G.\ 2000, ApJL, submitted

\bibitem[Bower 1997]{bower97} Bower, R.\ 1997, MNRAS, 288, 355

\bibitem[Cash 1979]{cash79} Cash, W.\ 1979, ApJ, 228, 939

\bibitem[Crawford \& Fabian 1996]{crawfab96} Crawford, C.S.\ \&
Fabian, A.\ 1996, MNRAS 282, 1483

\bibitem[Della Ceca et al.\ 2000]{dell00} Della Ceca, R., Scaramella,
R., Gioia, I.M., Rosati, P., Fiore, F., \& Squires, G.\ A\&A, 353, 498

\bibitem[Dickey \& Lockman 1990]{dl90} Dickey, J.M.\ \& Lockman, F.J.\ 1990, ARAA, 
28, 215

\bibitem[Donahue et al.\  1999]{don99} Donahue, M., Voit, G.M.,
Scharf, C.A., Gioia, I.M., Mullis, C.R., Hughes, J.P., \& Stocke,
J.T.\ 1999, ApJ, 527, 525

\bibitem[Edge \& Stewart 1991]{es91} Edge, A.C.\ \& Stewart, G.C.\
1991, MNRAS, 252, 428

\bibitem[Eke et al.\ 1998a]{eke98a} Eke, V., Cole, S., Frenk, C.
\& Henry, P.\ 1998, MNRAS, 298, 1145

\bibitem[Oukbir \& Blanchard 1992]{ob92} Oukbir, J.\ \& Blanchard, A.\ 
1992, A\&A, 262, L210

\bibitem[Raymond \& Smith 1977]{rs77} Raymond, J.C.\ \& Smith, B.W.\
1977, ApJS, 35, 419

\bibitem[Rosati et al.\ 1995]{rosati95} Rosati, P., della Ceca, R.,
Burg, R., Norman, C., \& Giacconi, R.\ 1995, ApJ, 445, L11

\bibitem[Rosati et al.\ 1998]{rosati98} Rosati, P., della Ceca, R.,
Norman, C., \& Giacconi, R.\ 1998, ApJ, 492, L21

\bibitem[Rosati et al.\ 1999]{rosati99} Rosati, P., Stanford, S. A.,
Eisenhardt, P., Elston, R., Spinrad, H., Stern, D., \& Dey, A.\ 1999, AJ, 118, 76

\bibitem[Schlegel, Finkbeiner, \& Davis 1998]{schlegel98} Schlegel,
D.J., Finkbeiner, D.P., Davis, M.\ 1998, ApJ, 500, 525

\bibitem[Stanford et al.\ 1997]{sas97} Stanford, S. A., Elston, R.,
Eisenhardt, P., Spinrad, H., Stern, D., \& Dey, A.\ 1997, AJ, 114, 2232

\bibitem[Tozzi \& Norman 2000]{tn00} Tozzi, P.\ \& Norman, C.\ 2000,
ApJ, in press

\end{thebibliography}
\end{document}